\documentstyle[aps,preprint]{revtex}
\def\laq{\ \raise 0.4ex\hbox{$<$}\kern -0.8em\lower 0.62
ex\hbox{$\sim$}\ }
\def\gaq{\ \raise 0.4ex\hbox{$>$}\kern -0.7em\lower 0.62
ex\hbox{$\sim$}\ }

\newcommand{\bb}{\bibitem}
\begin{document}

\preprint{\vbox{\baselineskip=12pt
\rightline{CERN-TH/99-381}
\vskip0.2truecm
\rightline{hep-th/9912055} \vskip1truecm}}

\vskip 2 cm

\title{A Causal Entropy Bound}

\author{R. Brustein${}^{1}$ and
G. Veneziano${}^{2}$}
\vskip 2 cm
\address{(1) Department of Physics,
Ben-Gurion University,
Beer-Sheva 84105, Israel \\
(2) Theory Division, CERN, CH-1211, Geneva 23, Switzerland}


\vskip 2 cm
\maketitle
\begin{abstract}

The identification of a causal-connection scale motivates us to
propose a new  covariant bound on entropy within a generic
space-like region. This  ``causal entropy bound", scaling as
$\sqrt{EV}$, and thus lying  around the geometric mean of
Bekenstein's $S/ER$ and holographic $S/A$ bounds,  is checked  in
various ``critical" situations. In the case of limited gravity,
Bekenstein's bound is the strongest while naive holography is the
weakest. In the case of strong gravity, our bound and Bousso's
holographic bound are stronger than Bekenstein's, while naive
holography is too tight, and hence typically wrong.
\end{abstract}

\vskip 1 cm

The second law of thermodynamics states that the entropy of a
closed system tends to grow towards its largest possible  value.
But what is this  maximal value? Bekenstein \cite{Bek1} has
suggested that, for a  limited gravity system of energy $E$, whose
size $R$ is larger than its gravitational radius, $R > R_g \equiv
2G_N E$,
  entropy is bounded by $S_{BEB}$,
\begin{equation}
 S_{BEB} = ER/\hbar = R_g~ R ~l_P^{-2},
\label{BEB}
\end{equation}
where $l_P$ is the Planck length (throughout this paper we will
 stress functional dependence, while
ignoring numerical factors, and set $c = k_B =1$).
 Note that (\ref{BEB}) bounds the
ratio  entropy/energy for a system of given size. The entropy of
normal systems, such as matter in thermal equilibrium, is well
below  bound (\ref{BEB}), moreover, in the 18 years which elapsed
since Bekenstein's proposal, and despite an ongoing debate
\cite{Bek2}, no physical example in which (\ref{BEB}) is violated
has been produced. Is the same bound applicable to more general
situations, for example, in cosmology? This question was addressed
by Bekenstein himself \cite{Bek3},
 who gave a prescription for a cosmological extension
by choosing $R$ in Eq. (\ref{BEB}) to be the particle horizon. Is
this a correct extension? And, even if so, is it possible to find
stronger
 bounds for systems that are not of limited gravity?

Holography \cite{holography}  suggests that maximal entropy is
bounded by $S_{HOL}$,
\begin{equation}
S_{HOL}= A l_P^{-2}, \label{HOL}
\end{equation}
where $A$ is the area of the space-like
surface enclosing a certain region of space.
For systems of limited gravity, since $R > R_g, A=R^2$,
 (\ref{BEB}) implies the holography bound (\ref{HOL}).
Is it possible to  extend entropy/area holographic  bounds to more
general situations, for example, to cosmology where, for large
enough regions, it soon becomes tighter than entropy/energy
bounds? This issue was first addressed by Fischler and Susskind
(FS) \cite{FS}, who proposed that the area of the particle
horizon should bound the entropy on the backward-looking
light cone according to (\ref{HOL}). It was soon realized, however, that
the FS proposal requires modifications, since violations of it were
found to occur in physically reasonable situations.

  Several attempts were made
to mend the FS proposal, first within cosmology
 \cite{EL,GV1,BR,KL}, and then, by Bousso,
 in arbitrary space-times \cite{Bousso}.
In some cosmological situations, Bousso's proposal  reduces to the
previously proposed ones, which identify the maximal size of a
spatial region for which holography works with the Hubble radius
\cite{EL}, \cite{GV1} (or apparent horizon \cite{BR}), while in
other situations it is quite different. The advantages of Bousso's
proposal are that $(i)$ it is very general; and $(ii)$ it is
manifestly covariant. A possible shortcoming of the proposal is
that it bounds entropy on light-like hypersurfaces:
 in order  to extend the bound to  space-like regions,
a ``space-like projection",
 which is not always  possible, has to be performed \cite{Bousso}.

We make here a motivated proposal for an improved covariant
bound, applicable to  entropy on space-like hypersurfaces, and
test it in several critical cases. We then
 compare our bound to other proposals, in particular
to Bekenstein's and  Bousso's, and show that, for systems of
limited gravity, Bekenstein's bound is the tightest, while, in other
situations, our bound is the strongest one  proposed so far
which does not lead to contradictions for space-like regions.
A crucial difference between our proposal and Bousso's is that
Bousso decides from the start to look for a holographic
 $S/A$  bound and, as he points out, this forces him to consider
light-like hypersurfaces. We do not insist, a priori, on
a holographic bound, but  aim at generality of the hypersurface
and check how holography may or may not work a posteriori.

Let us first state our proposal, and then motivate and test it. Consider a
generic spacelike hypersurface, defined by the equation $\tau =
0$, and a compact region lying within it defined by $\sigma \le
0$. We propose that the entropy contained in this region,
$S(\tau = 0, \; \sigma \le 0)$, is
bounded  by $S_{CEB}$, 
\begin{eqnarray}
&&S_{CEB}= l_P^{-2} \int\limits_{\sigma <0} d^4 x \sqrt{-g}
 \delta(\tau)
 \sqrt{ {\rm Max}_\pm\left[ (G_{\mu\nu}\pm R_{\mu\nu})
\partial^{\mu} \tau \partial^{\nu} \tau \right]}~ =
 \nonumber \\
&& l_P^{-1} \hbar^{-1/2} \int\limits_{\sigma <0} d^4 x \sqrt{-g}
 \delta(\tau)
\sqrt{ {\rm Max}_\pm \left[( T_{\mu\nu} \pm T_{\mu\nu} \mp {1\over2}
g_{\mu\nu}~T)
\partial^{\mu} \tau \partial^{\nu} \tau \right]}.
\label{CCB}
\end{eqnarray}
Here $G_{\mu\nu}$,  $R_{\mu\nu}$ are the Einstein and Ricci
tensor, respectively, $T_{\mu\nu}$ is the energy-momentum tensor,
and $T$ its trace. To derive the second equality  we have used
Einstein's equations, $G_{\mu\nu}  = 8 \pi G_N T_{\mu\nu}$. Note
the appearance of the square-root of the energy contained in the
region,  which we alluded to in the abstract. Note also that
(\ref{CCB}) is manifestly covariant, and invariant under
reparametrization of the hypersurface equation:  such an
invariance requires a square-root of $\partial^{\mu} \tau
\partial^{\nu} \tau $. Reality of $S_{CEB}$ is assured if sources
obey the weak energy condition, $T_{\mu\nu}\partial^{\mu} \tau
\partial^{\nu} \tau \ge 0$, since then the sum of the two
combinations in (\ref{CCB}), and thus their maximum, are positive.
In the limit in which the hypersurface is lightlike,
Eq.(\ref{CCB}) becomes:
\begin{eqnarray}
&&S_{CEB}= \int_{\sigma <0} d^4 x \sqrt{-g}
 \delta(\tau)
 \sqrt{R_{\mu\nu}
\partial^{\mu} \tau \partial^{\nu} \tau}
 \nonumber \\
&& =  \int_{\sigma <0} d^4 x \sqrt{-g}
 \delta(\tau)
\sqrt{T_{\mu\nu}
\partial^{\mu} \tau \partial^{\nu} \tau} \;,
\partial^{\mu} \tau \partial_{\mu} \tau = 0 \;,
\label{CCBnull}
\end{eqnarray}
and appears to be related to one of the assumptions  made in
\cite{Wald} (Eq. (1.10)) in order
to derive a modified Bousso-type bound. We should stress, however,
that (\ref{CCBnull}) does not necessarily follow from the
arguments given below in support of (\ref{CCB}).

The physical motivations leading us to the
above proposal are similar to those
 used in the recently proposed Hubble-entropy-bound (HEB)
 \cite{GV1}
(see also \cite{EL,BR,KL}) i.e.:
$(i)$ that entropy is maximized, in a given region of space, by
the largest black hole that can fit in it $(ii)$ that the largest
black hole that can hold together without falling apart in a
cosmological background has typically the size of the Hubble horizon.
 This second, crucial assumption appears to be
supported qualitatively
by a number of previous results \cite{Carr}, but clearly needs
 to be refined and, possibly, to be defined covariantly.
With such a goal in mind, we will proceed as follows: we will
start by identifying a critical (``Jeans") length scale above
which perturbations are causally disconnected so that black holes
of larger size, very likely, cannot form.  We will first find this causal
connection (CC) scale $R_{CC}$ for the simplest cosmological
backgrounds, then extend it to more general cases and, finally,
 guess the completely general expression using general covariance.

In order to identify the CC scale for a homogeneous, isotropic,
and spatially flat background, let us consider a generic
perturbation around such a background in the hamiltonian approach
developed in \cite{BMV}. The Fourier components of the
(normalized)  perturbation and of its (normalized) conjugate
momentum satisfy Schroedinger-like equations
${\widehat\Psi}_k{''}\!\!+\!\!\left[k^2-(S^{1/2}){''}
S^{-1/2}\right]{\widehat\Psi}_k\!\!\!=\!\!\!0$,
${\widehat\Pi}_k{''}\!\!+\!\! \left[k^2-(S^{-1/2}){''}
S^{1/2}\right] {\widehat\Pi}_k\!\!\!=\!\!\!0$, where $k$ is the
comoving momentum, a prime denotes differentiation w.r.t.
conformal time $\eta$, and $S^{1/2}$ is the so-called ``pump
field", a combination of the various backgrounds which depends
on the specific perturbation under study.
The perturbation equations clearly identify a ``Jeans-like" CC
comoving momentum
\begin{eqnarray}
  k_{CC}^2  &=&  {\rm Max} \left[ (S^{1/2}){''}
S^{-1/2} ~, ~(S^{-1/2}){''}
S^{1/2} ~ \right]
  \nonumber \\ &=&
{\rm Max} \left[ {\cal K}' + {\cal K}^2 ~,
~ - {\cal K}' + {\cal K}^2 \right],
\label{kjeans}
\end{eqnarray}
where ${\cal K}=(S^{1/2}){'} S^{-1/2}$.
 Note that eq. (\ref{kjeans}) always defines
a real $k_{CC}$ since the sum of the two quantities appearing on
the r.h.s. is positive semidefinite.
 Since tensor perturbations are always present, let us
restrict our attention to them. The ``pump
field" $S^{1/2}$ is simply given, in this case,
 by the scale factor $a(\eta)$ so that
${\cal K} \rightarrow {\cal H} = a'/a$.
 Eq. (\ref{kjeans}) is
immediately converted into the definition of a proper ``Jeans" CC
length $R_{CC} = a k_{CC}^{-1}$. Substituting into
eq.(\ref{kjeans}), and expressing the result in terms of
proper-time quantities, we obtain (for tensor perturbations)
\begin{equation}
 R_{CC}^{-2} = {\rm Max}
 \left[ \dot{H} + 2 H^2~, ~ - \dot{H}~\right].
\label{RCC}
\end{equation}

Before trying to recast eq.(\ref{RCC}) in a more covariant form
let us remove the assumption of spatial flatness by introducing
the usual spatial-curvature parameter $\kappa$ ($\kappa = 0,
\pm1$). The study of perturbations in non-flat space
\cite{Garriga} is considerably more complicated than  in a
spatially-flat background. The final result, however, appears to
be extremely simple \cite{GPV},  and
 can be obtained from the flat case by the following replacements
 in eq.(\ref{kjeans}):
${\cal H}^2 \rightarrow {\cal H}^2 + \kappa$,
${\cal H}' \rightarrow  {\cal H}' $.
Using this simple rule
(see below for another confirmation of its validity) we
arrive at the following generalization of (\ref{RCC})
\begin{equation}
  R_{CC}^{-2} = {\rm Max} \left[ \dot{H} + 2 H^2~ + \kappa/a^2, ~
 - \dot{H}~ + \kappa/a^2 \right].
\label{RCCk}
\end{equation}

At this point we could have introduced anisotropy in our
homogeneous background and study  perturbations  with or
without spatial curvature. This should certainly be done
as a check of a short-cut procedure based on general
covariance that we adopt instead.
We note that the $00$ components of the
Ricci and Einstein tensors for our background are given by
\begin{equation}
  R_{00} = -3 (\dot{H} +  H^2) ~, ~ G_{00} = 3 ( H^2 + \kappa/a^2)\;.
\label{RandG}
\end{equation}
Obviously,
\begin{eqnarray}
  R_{CC}^{-2} &=& {1\over 3}{\rm Max}_{\mp}
  \left(G_{00} \mp  R_{00} \right) \nonumber \\  &=& 4 \pi G_N
~{\rm Max} \left[{\rho \over 3} - p~,~ \rho + p \right] \;,
\label{RCCC}
\end{eqnarray}
where we have inserted Einstein's equations using, as an example,
a perfect-fluid energy momentum tensor,
$T^\mu_\nu=diag(\rho,-p,-p,-p)$. Eq.(\ref{RCCC}) is guaranteed to
define a real $R_{CC}$ if the weak energy condition
(reading here
 $\rho
>0$)  holds, since the sum of the two combinations is positive in this
case \cite{com1}.

As a final step, let us convert eq.(\ref{RCCC}) into an explicitly
covariant bound on entropy using, as in \cite{GV1}, the idea that
entropy is maximized by having  maximal size black holes filling
up the volume. Using $R_{CC}$ as the maximal scale for black
holes, we get a bound on entropy which scales like
$
  S \sim V R_{CC}^{-3} ~ R_{CC}^2 l_P^{-2}~
  = ~ V R_{CC}^{-1} ~  l_P^{-2}.
$
We now express $R_{CC}^{-1}$ as in (\ref{RCCC}) in terms of the
components of the Ricci and Einstein tensors in the direction
orthogonal to the hypersurface on which the entropy is being
computed. This can be done covariantly by defining the
hypersurface through the equation $\tau =0$ and by identifying the
normal with the vector $\nabla^{\mu} \tau$. This procedure leads
immediately to the proposal (\ref{CCB}).

  Alternatively, using
standard $3+1$ ADM formalism \cite{ADM}, we can express the
relevant components of the Ricci and Einstein tensors in terms of
the intrinsic and extrinsic curvature of the hypersurface under
study and arrive at the following final formula:
\begin{eqnarray}
 S_{CEB} &=&  l_P^{-2}
 ~\int d^3 x \sqrt{h}   \;
\left[{\rm Max}\left(P~,~Q \right)\right]^{1/2}~, \label{CCBADM}
\end{eqnarray}
 where $P = {1 \over 2} {\cal R} +
\dot{\theta}
 +{2 \over 3} \theta^2 + \sigma^2 -
{\cal A}~, Q = {1 \over 2} {\cal R} - \dot{\theta} - 3 \sigma^2 +
{\cal A}$, and using  standard notations,
 we have denoted by ${\cal R}$ the intrinsic 3-curvature scalar,
  by $\theta$ the expansion
 rate, by $\sigma$ the shear, and by ${\cal A}$ the ``acceleration" given
 (for vanishing shifts $N_i$)
 in terms of the
 lapse function $N$ by ${\cal A} = N^{-1} N^{,i}_{\ ;i}$.

We turn to check our proposal for various physical systems and to
verify that it is sensible.

{\em 1. Systems of limited gravity}

 \noindent
We note (see below) that for systems of limited gravity the
Bekenstein bound is tighter than ours $S_{BEB}<S_{CEB}$. Therefore,
in all systems for which the BEB is obeyed,  ours will be obeyed as well.

{\em 2. Cosmology}

 \noindent
The universe is a system of strong self-gravity. The geometry of
the universe is determined by self-gravity, and the size of the
universe is at least its gravitational radius. The strongest
challenges to entropy bounds in general, and to our bound in
particular, come from considering (re)collapsing universes. We
discuss three cases basically exhausting the possible types of
cosmologies:
\begin{itemize}
\item {(i)} $|\dot H|\sim H^2$. In this region effective energy density
and pressure are of the same order, $\rho\sim p$, and all previously
suggested length scales that should be considered in entropy
bounds, such as particle horizon, apparent horizon, $R_{CC}$,
Hubble length,  are parametrically equal. In particular, it is
already established that HEB is not violated if some reasonable
restrictions on the equation of state are imposed, and therefore our
bound (and others) is also valid.
\item {(ii)}  $|\dot H|\ll H^2$. In
this case $|\rho+ p| \ll \rho$, and the universe is
inflationary. Here the naive holographic bound fails
miserably, but HEB does well. Since in this case $R_{CC}$ is
parametrically equal to $|H|^{-1}$, it follows that CEB works as
well as HEB.
\item (iii) $|\dot H|\gg H^2$ i.e. $|\rho|
\ll p$. Since $\rho$ and $p$ are the effective energy density and
pressure, there are no problems with causality. This case occurs,
for instance, near the turning point of an expanding universe
which recollapses as the result of a negative cosmological
constant, of positive spatial curvature (or of both).
 Both the naive $HEB$ and the apparent horizon bound (AHB) of
\cite{BR} fail here, while Bousso's prescription  does fine. We
would like to show that CEB can easily cope with this third case.
\end{itemize}

Consider  either a  flat or  closed
universe  with some perfect fluid in thermal equilibrium and a
constant equation of state $p=\gamma \rho~, 1>\gamma > -1$,
and with an additional small negative cosmological constant
$\Lambda = - \lambda$.
The universe starts out expanding, reaches a maximal size, and
then contracts towards a singularity.  In this case, matter entropy
within a comoving volume is constant in time, but near the point
of maximal expansion the apparent horizon, and the Hubble length,
diverge, causing violation of the HEB and AHB. However, for a fixed
comoving volume, $S_{CEB}\sim a^3
R_{CC}^{-1}$, and, since $R_{CC}$ is never larger than some maximal
value, CEB has a chance of doing better.

To see this explicitly, note that, in this case, and independently
of $\kappa$, $R_{CC}^{-2} = 1/3 {\rm Max} [ 1/2 \rho_0 (1-3\gamma)
a^{-3(1+\gamma)}- 2 \lambda~,~~ 3/2 (1 + \gamma)  \rho_0
a^{-3(1+\gamma)}] \ge 1/2 (1 + \gamma)
 \rho_0 a^{-3(1+\gamma)}$, where $\rho_0$ is the initial
energy density and $a$ is the ratio of the scale factor
to its initial value. It follows that, in a fixed comoving volume,
$S_{CEB} \sim  a^{3/2(1-\gamma)}$. Since $\gamma < 1$, this means that
$S_{CEB}$ grows during the expansion, reaches a maximum at the turning
point, and then starts decreasing. If we give initial conditions at a time when
curvature and cosmological constant are negligible, which is always the case
at sufficiently early times, CEB will be obeyed initially
provided energy density and curvature are less than Planckian. But then
the above evolution of $S_{CEB}$ will guarantee that the bound is fulfilled
at all times until Planckian density and curvature is reached in the
recollapsing phase, i.e. throughout the classical evolution of our Universe.

In spite of the above encouraging results, we would like to
express caution about the assumption of homogeneity and isotropy.
Since we are considering matter in thermal equilibrium at
temperature $T$, it has mass fluctuations
 $\left(\frac{\delta M}{M}\right)_\ell\sim (\ell T)^{-3}$,
in a region of size $\ell$, which, once the universe starts
contracting, begin to grow on time scales $R_{CC}^{-1}$, so
assuming homogeneity and isotropy could be incorrect.

{\em 3. Collapsing regions}

 \noindent
In this case we have limited computational power. We can
qualitatively check cases that are similar to the cosmological
ones \cite{MTW2}, such as  homogeneous, isotropic contracting
pressureless regions, or a contracting homogeneous, isotropic
region filled with a perfect fluid. The pressureless case can be
described by a Friedman interior and a Schwarzschild exterior.
Since CEB is valid for the analogue cosmological solution it is
also valid for this case.

A particularly interesting case
is that of the (generically non-homogeneous)
 collapse of a stiff fluid ($p = \rho$), which,
up to a simple field redefinition, can be mapped onto the
dilaton-driven inflation of string cosmology \cite{BDV}. In this
case one finds a constant $S_{CEB}$ in agreement with the HEB
result \cite{GV1}. Hence, no problem arises in this case, even if
one starts from a saturated $S_{CEB}$ at the onset of collapse.
For non-stiff  equations of state, the situation appears less safe
if one starts near saturation (this was already pointed out in
\cite{Bousso} for Bousso's case). As Bousso himself points out,
however, care must be taken in this case of  perturbations which
tend to grow non linear and form singularities on rather short
time scales. Such cases  cannot be described analytically, but
have been looked at numerically.
 We believe that numerical investigations of collapsing
systems will be extremely useful in determining the general
validity of CEB and of other entropy bounds.

Finally, we compare  our CEB to previously proposed bounds, in
particular to Bekenstein's and  Bousso's. For systems of limited
gravity whose size  exceeds their Schwarzschild radius: $R > R_g$,
Bekenstein's bound is given by $S<S_{BEB} = l_P^{-2} R~R_g$, and
Bousso's procedure results in the holography bound (\ref{HOL})
\cite{Bousso},
 $S < S_{HOL} =  l_P^{-2} R^2$, but since $R > R_g$,
 $S_{BEB}<S_{HOL}$,
and therefore  Bousso's bound  is less stringent than
Bekenstein's. Consider now  CEB applied to the region of size $R$
containing an isolated system. Expressing CEB in the form
(\ref{CCB}) one immediately obtains:
\begin{equation}
  S_{CEB} =  l_P^{-1} R^{3/2} E^{1/2} \hbar^{-1/2}
 = (S_{HOL}~~S_{BEB})^{1/2}\; ,
 \label{CCBB}
\end{equation}
implying
 $$
S_{BEB}\le S_{CEB} \le S_{HOL}.
 $$
We conclude that for isolated systems of limited self-gravity the
Bekenstein bound is the tightest, followed by our CEB and,
finally, by Bousso's holographic bound. 
Similar scaling properties for the HEB were discussed in \cite{GV1}. 
The same scaling laws also follow from (apparently unrelated) 
quantum measurement arguments, see \cite{sasakura}.

For regions of space that contain so much energy that the
corresponding gravitational radius $R_g$ exceeds $R$, Bekenstein's
bound is the weakest, while the naive holography bound is the
strongest (but very often wrong).  Bousso's proposal (see also
\cite{BR})  uses the apparent horizon $R_{AH}$ while CEB uses
$R_{CC}$. For homogeneous cosmologies, $R_{CC}<R_{AH}$, since
 $R_{CC}^{-2}$, according to (\ref{RCCk}), is always larger than
the average of the two terms appearing on its r.h.s., which is precisely
 $R_{AH}^{-2} =  H^2+\kappa/a^2$.
 Since, for a fixed volume, the bounds scale like $R_{AH}^{-1}$ or
$R_{CC}^{-1}$, we immediately find that CEB is generally more generous
 than the AHB
proposed in \cite{BR}. This is what makes it possible for CEB to
be fulfilled in the negative cosmological case discussed above,
where AHB or HEB are violated. Comparison with Bousso's proposal
is more subtle, since he makes use of the AH area to bound entropy
on light sheets. This can be converted into a bound on the entropy
of the space-like region inside the AH only in special cases. That
is what makes Bousso's bound fulfilled even when the straight AHB
fails.

 To summarize, we have found that, for systems
of strong gravity,
 $$
 S_{AHB} \sim S_{HEB} \le { S_{CEB}, S_{Bousso}} \le  S_{BEB}\; ,
 $$
i.e., that our CEB and
  Bousso's holographic bound are the strongest (yet apparently safe) bounds,
while Bekenstein's is the weakest. Instead, the naive holography
bound comes out badly: implied by Bekenstein's in the case of
limited gravity, and too tight (and hence typically wrong) in the
case of strong gravity. Bousso's modification fares much better
than the naive holographic bound.  We believe, however, that CEB,
by being applicable to space-like regions, stands out in terms of
its physical motivations and  its potential for practical uses.

We anticipate applications of CEB to the study of the possibility
that black hole remnants carry enough entropy to restore unitary
evolution.  CEB may be converted into a new kind of generalized
second law following  \cite{RB}, and used to study cosmological
singularities. The investigation of these possibilities is left,
however, to future work.

\acknowledgements
 We thank J. Bekenstein, R. Bousso, M. Gasperini,
N. Kaloper, G. Kane, A. Linde and R. Madden for comments,
suggestions, criticism, and useful discussions, and N. Sasakura
for informing us about his work. R.B. would like to
thank the theory division at CERN for hospitality.

\end{document}